\documentclass[trackchanges,twocolumn]{aastex7}

\usepackage{amsmath}
\usepackage{svg}
\usepackage{subcaption}
\usepackage{comment}
\usepackage{booktabs}
\usepackage{multirow}
\usepackage{soul}

\makeatletter
\renewcommand{\frontmatter@title@above}{}
\makeatother

\begin{document}

\title{The Role of Non-local Thermal Transport in Flare-Driven Chromospheric Evaporation}

\author[orcid=0009-0008-2585-9280, gname=Thomas, sname='Parmenter']{T. Parmenter} 
\affiliation{Centre for Fusion, Space and Astrophysics, Department of Physics, University of Warwick,  Coventry CV4 7AL, UK}
\email{thomas.parmenter@warwick.ac.uk}

\author[orcid=0000-0002-9322-4913,
gname=Tony, sname='Arber']{T. Arber} 
\affiliation{Centre for Fusion, Space and Astrophysics, Department of Physics, University of Warwick,  Coventry CV4 7AL, UK}
\email{T.D.Arber@warwick.ac.uk}

\author[orcid=0000-0002-3505-9542
,gname=Sergey, sname='Belov']{S. Belov} 
\affiliation{Centre for Fusion, Space and Astrophysics, Department of Physics, University of Warwick,  Coventry CV4 7AL, UK}
\email{Sergey.Belov@warwick.ac.uk}

\author[orcid=0000-0003-0784-1294
,gname=Tom, sname='Goffrey']{T. Goffrey} 
\affiliation{Centre for Fusion, Space and Astrophysics, Department of Physics, University of Warwick,  Coventry CV4 7AL, UK}
\email{T.Goffrey.1@warwick.ac.uk}

\collaboration{all}{}

\begin{abstract}

Thermal conduction plays an important role in energy transport during solar flares. Flares heat the solar atmosphere to tens of MK which, combined with mean free paths in excess of 1Mm, can cause the heat flux to become non-local. We use the magnetohydrodynamic Lare2d code to simulate a reconnective heating event with three different thermal conduction models: classical Spitzer-H\"{a}rm (SH) along the magnetic field, local transport with a heat flux limiter (FL), and the non-local Schurtz, Nicola$\ddot{\mathrm{i}}$, Busquet (SNB) model which is routinely used in laser-plasma codes and has been benchmarked against full Vlasov-Fokker-Planck treatment. The evolution of the atmosphere post heating is compared between simulations using each of these models. Due to a supressed heat flux, SNB gives rise to significantly higher peak temperatures than SH. Further, with SNB transport chromospheric evaporation occurs later and has stronger extreme ultraviolet emission intensity. FL simulations have similar temperature evolution and chromospheric evaporation to SNB, however different values of the flux limiter work best for peak temperature, downflow velocity and EUV emission intensity, and so there is no one optimal value for all cases. We conclude that the SNB model provides a practicable solution to incorporate non-local thermal transport at MHD scales, which will be crucial to improve the realism and accuracy of the next generation of solar flare simulations.

\end{abstract}


\section{Introduction} 

Solar flares are some of the most energetic events in the solar system, releasing up to $10^{25}$J of energy stored in the Sun's magnetic field during X-class flares and producing radiation across the electromagnetic spectrum \citep{Benz2016}. The energy is released rapidly: over the course of seconds to minutes, and is localised to flare ribbons less than 1Mm across, giving rise to extremely high temperatures, reaching well in excess of 10MK in large flares \citep{Doschek1987, Harra2023, daSilva2023, Russell2025}. Combined with the low density found within the solar atmosphere, this can lead to the thermal electron mean free path $\lambda_T$, which gives a characteristic length scale for collisions of electrons moving at the thermal speed, becoming comparable in size to the characteristic temperature length scale $L_T = T / \left|\nabla T\right|$. The ratio between these two length scales is quantified using the Knudsen number $K_n = \lambda_T / L_T$. Classically thermal transport in a plasma is described using the local model described by \cite{Spitzer1953}, however it has been shown that when $K_n > 0.01$ \citep{Arber2023} or possibly for even lower values \citep{Belov2025_2}, thermal transport becomes non-local and so the classical Spitzer-H\"{a}rm (SH) model is no longer valid. Non-local thermal transport is characterised by a reduced bulk heat flux as well as preheating, where the spatial profile of the heat flux becomes broader due to transport by the higher-energy electrons with longer mean free paths. 

To date, there is observational evidence for the suppression of thermal conduction in hot flaring loops. In particular, \mbox{\citet{Battaglia2009}} analysed pre-flare emission observed by the Reuven Ramaty High Energy Solar Spectroscopic Imager (RHESSI) and detected an increase in the emission measure and plasma density. This increase was interpreted as chromospheric evaporation driven by a flux-limited conductive heat flux. Moreover, $K_n$ estimates showed that classical Spitzer conduction is not applicable under the inferred flare conditions, with the conductive flux being locally limited or potentially reaching the non-local saturated regime. Later, \mbox{\citet{Imada2015}} observed a strong redshift at the base of a flaring loop using the Hinode/EUV Imaging Spectrometer. Their subsequent 1D hydrodynamic simulations reproduced the observed redshift only when the thermal conductivity was reduced to 10\% of its classical value, providing possible evidence for thermal-conduction suppression during the impulsive phase of flares. Moreover, \mbox{\citet{Wang2015}} used a standing slow-mode wave to seismologically infer the polytropic index in a $\sim10$\,MK flaring loop. The measured value, 1.64, is close to the adiabatic index of an ideal monatomic gas, which was interpreted as evidence for an approximately threefold suppression of thermal conduction.

 Despite this, there have only been a limited number of studies within solar physics that model thermal conduction non-locally \citep{Karpen1987, Silva2018, Belov2025}. The most accurate way to model non-local energy transport is to use the full Vlasov-Fokker-Planck (VFP) treatment, for example \mbox{\citet{Zharkova2011}}. However, on the temporal and spatial scales of solar flare simulations using MHD in 2D or 3D this is not computationally feasible, and so models which can be more easily implemented into an MHD code must be used. 
 A common approach taken within MHD simulations, e.g. by \citep{Cheung2019, Gonzalez2024, Johnston2017}, is to use classical thermal conduction but place a limit on the maximum value of the heat flux. This reduces the bulk heat flux but fails to replicate the preheating of a proper non-local model. Unless in direct comparison to observations or a non-local model, it also is reliant on correctly choosing the exact value of the maximum flux, which is not universal \citep{Jones2017, VanderHolst2014}.

 There have been some attempts at using a fully non-local treatment for the heat flux in solar physics, utilising the \textbf{LMV} model \citep{Luciani1983}. \cite{Karpen1987} simulated a flaring atmosphere using a 1D model and compared results between classical, locally limited, and non-local thermal transport. They found higher temperatures were reached in the solar atmosphere when using a non-local model and preheating of the chromosphere lessened the transition region temperature gradient. Downward velocities were initially higher from the flaring site, whereas lower upward velocities were seen during chromospheric evaporation. \cite{Silva2018} carried out 3D simulations of a non-flaring active region of the solar atmosphere. They found the most significant difference between the non-local treatment and classical transport in the transition region due to the high temperature gradients. Non-local transport resulted in a smoother temperature gradient within the transition region as well as shifting it upwards approximately 500km. 
 
 The LMV model is fundamentally 1-dimensional in its formulation, convolving the local heat flux along the magnetic field with a kernel. In order to incorporate the LMV model into 3 dimensions, \cite{Silva2018} were required to trace field lines and then calculate thermal conduction along each field line which greatly increases the computational cost. For this reason, within laser-plasma physics, the $\textbf{SNB}$ model \citep{Schurtz2000} has become the most effective way to treat non-local thermal conduction at the MHD scale, superseding the LMV model. The SNB model requires solving differential equations without any tracing of magnetic field lines which makes its implementation into codes easier. It has been shown to have good agreement with full VFP simulations \citep{Marocchino2013, Brodrick2017} and requires no free parameters, and hence is used regularly in laser-plasma codes \citep{Chen2023, Lau2025}.  
 
It has been shown, with 1D solar flare simulations using the SNB model, that non-local transport can strongly influence the evolution of the density and temperature at the apex of the coronal loop, resulting in higher temperatures and lower densities during the loop evolution \citep{Belov2025}. In particular, heat-flux suppression at the energy-deposition stage creates more localised and intense temperature peaks compared to those produced by local models. The lower amount of energy reaching the chromosphere leads to reduced chromospheric evaporations and, therefore, lower post-flare densities. In this paper, we investigate the influence of non-local transport effects on the coronal and chromospheric plasma response to magnetic energy release by extending the 1D study of \citet{Belov2025} to a 2D flare model. This extension is essential because a 2D model self-consistently captures magnetic reconnection, plasma outflows, and MHD waves, allowing the magnetic energy, released via reconnection, to be partitioned between thermal energy, bulk flows, and waves. Consequently, the amount of energy available for thermal conduction to the chromosphere may differ significantly from that predicted by 1D models, potentially leading to a different chromospheric evaporation response.

The paper is organised as follows. Section \ref{sec:model} of this paper outlines the methodology, providing the governing equations and models used within the simulation. Section \ref{sec:results} describes and discusses the results obtained, looking at the temperature evolution and chromospheric response following reconnective heating. Conclusions are presented in Section \ref{sec:conc}. 

\section{Models} \label{sec:model}

\subsection{Governing Equations}
Simulations are carried out using the Lare2d code \footnote{github.com/Warwick-Plasma/Lare2d} to solve the MHD equations in 2.5D, i.e. with vectors having three components but quantities only varying on a 2D grid \citep{Arber2001}. The simulation domain is rectangular with $-20 < x < 20$ and $0 < y < 60$ in Mm. A resolution of $800 \times 1200$ grid cells is used, giving a uniform grid spacing of 50km. The code uses a Lagrangian-Eulerian remap scheme to solve the following equations expressed in Lagrangian form:  
\begin{equation}
\frac{\mathrm{D}\rho}{\mathrm{D}t} = - \rho \nabla.\textbf{v},
\end{equation}

\begin{equation}
\frac{\mathrm{D}\textbf{v}}{\mathrm{D}t} = \frac{1}{\rho}\left(\nabla \times \textbf{B} \right) \times \textbf{B} - \frac{1}{\rho} \nabla P - \textbf{g},
\end{equation}

\begin{equation}
\frac{\mathrm{D}\textbf{B}}{\mathrm{D}t} = \left(\textbf{B} .\cdot\nabla \right)\textbf{v} - \textbf{B}\left( \nabla \cdot \textbf{v} \right) - \nabla \times \left(\eta \nabla \times \textbf{B} \right),
\end{equation}
\begin{equation}
\frac{\mathrm{D}\epsilon}{\mathrm{D}t} = -\frac{P}{\rho} \nabla.\textbf{v} + \frac{\eta}{\rho} \left|\textbf{j}\right|^2 - \nabla \cdot \textbf{q} -n_e^2 \Lambda\left(T\right) + H,
\end{equation}

where $\rho$ is the mass density, \textbf{v} is the velocity, \textbf{B} is the magnetic field strength, $P$ is the gas pressure, $\textbf{g}=-274 \ \hat{\textbf{y}} \ \mathrm{ms^{-2}}$ is the gravitational acceleration, $\eta$ is the resistivity, \textbf{j} is the current density, $\epsilon$ is the specific internal energy density satisfying $P = \rho \epsilon \left(\gamma-1\right)$ with $\gamma=5/3$ the ratio of the specific heats. $n_e$ is the electron number density, $\Lambda\left(T\right)$ is the radiative loss function for an optically thin plasma obtained using the CHIANTI database \citep{Dere1997,DelZanna2021}, $H$ is a uniform coronal heating term of $10^{-5}\,$Wm$^{-1}$ used to sustain the background coronal temperature and \textbf{q} is the heat flux. The plasma is assumed to be fully ionised with a mean ion mass of $1.2 m_p$, where $m_p$ is the mass of a proton.

\subsection{Thermal Conduction}

Three different models are compared in this paper for the heat flux $\textbf{q}$: 
\begin{enumerate}
\item{The classical model which uses the local Spitzer-H\"{a}rm approximation \citep{Spitzer1953} for thermal transport along the magnetic field. The total expression, including cross-field transport, for the heat flux is as follows:
\begin{equation}
\textbf{q}_{SH} = \frac{\kappa_0}{|\textbf{B}|^2 + B_{min}^2} \left[(\textbf{B} \cdot \nabla T) \textbf{B} + B_{min}^2 \nabla T\right],
\end{equation}
where $\kappa_0$ is $9.1 \times 10^{-12} \mathrm{Wm}^{-1}\mathrm{K}^{-7/2}$ and $B_{min}$ is a parameter to ensure that thermal conduction smoothly transitions from field-aligned to isotropic as $|\textbf{B}| \to 0$.  

Henceforth referred to as SH conduction.}
\item{Local thermal transport but with a limit on the maximum value of the heat flux: 
\begin{equation}
\textbf{q}_{FL} = \frac{\alpha q_{fs}}{\sqrt{\left(\alpha q_{fs}\right)^2 + |\textbf{q}_{SH}|^2}} \textbf{q}_{SH},
\end{equation}
where $q_{fs} = (n_e k_B T v_T)$ is the free streaming flux, i.e. if the thermal energy density was transported at the thermal speed $v_T$, and so gives a sensible upper bound on the heat flux. $\alpha$ is a free parameter in the range 0.01-1 within our simulations. This encapsulates the range of values used in other works including $\alpha=0.03$ in \cite{Jones2017}, $\alpha = 1 /6$ in \cite{Cheung2019} and $\alpha = 0.53$ in \cite{Karpen1987}. A default value of 0.2 is used unless stated otherwise. 

This model is referred to as flux-limited (FL) thermal conduction.}
\item{Non-local thermal transport using the model first described by \citep{Schurtz2000}. In the first step in this model, the heat flux $\textbf{q}_{SH}$ is calculated using the Spitzer-H\"{a}rm approximation. The final heat flux is calculated as the sum of the SH flux and a non-local correction. The electron energy distribution function is divided into $N_g$ energy groups to account for the fact that the size of the correction depends on the energy of the electrons transporting the flux. The maximum energy for each bin is $E_g$. It should be noted that calculating the correction adds an extra computational cost, increasing the runtime of the thermal conduction step by a factor of approximately the number of energy bins. In our code, 20 energy bins are used which are cubically spaced, so as to still correctly model lower energy regions, between 0K and 100 times the maximum temperature. The heat flux $\textbf{q}_{SH}$ is then divided up, proportionally to the contribution from each electron-energy group to the total heat flux, among these energy bins: 
\begin{equation}
\textbf{U}_g = \frac{1}{24} \int_{E_{g-1} / k_BT}^{E_g / k_BT} \beta^4 e^{-\beta} d\beta \ \textbf{q}_{SH}.
\end{equation} 
Then the following equations are solved to calculate the corrected heat flux:
\begin{equation}
\left(\frac{1}{\lambda_g} - \nabla \frac{\lambda_g}{3} \nabla \right) H_g = -\nabla \cdot \textbf{U}_g,
\end{equation}
\begin{equation}
\textbf{q}_{SNB} = \textbf{q}_{SH} - \sum_{g=1}^{N_g} \frac{\lambda_g}{3} \nabla H_g,
\end{equation}
where $\lambda_g$ is the average mean free path for each energy group, i.e. $\lambda_g = 2 \ (E_{g-\tfrac{1}{2}} / k_BT)^2\lambda_T$ with $\lambda_T = 6 \sqrt{2 \pi^3} \, {k_B}^2 {\epsilon_0}^2 / (e^4 \log{\Lambda}) \left[T^2 / n_e \right]$. 

This model is referred to as the SNB thermal conduction. 
}
\end{enumerate}

\subsection{Boundary Conditions and Atmosphere}

 The lower boundary at $y = 0$ is taken to be the solar surface with $\rho$ and $T$ fixed, $\textbf{v} = 0$ and $\partial \textbf{B}/ \partial y = \textbf{0}$. Zero gradient conditions are used for $\rho$, $T$, $\textbf{B}$, $v_y$ and $v_z$ at both the left and right boundaries, whereas $v_x$ is set to zero. The upper boundary is placed sufficiently far away that it has limited effect but zero gradient conditions are placed on $\rho$, $T$, $\textbf{v}$, $B_y$ and $B_z$, with $B_x$ taken to be asymmetric across the boundary.   

\begin{figure*}
\includegraphics[width = 180mm]{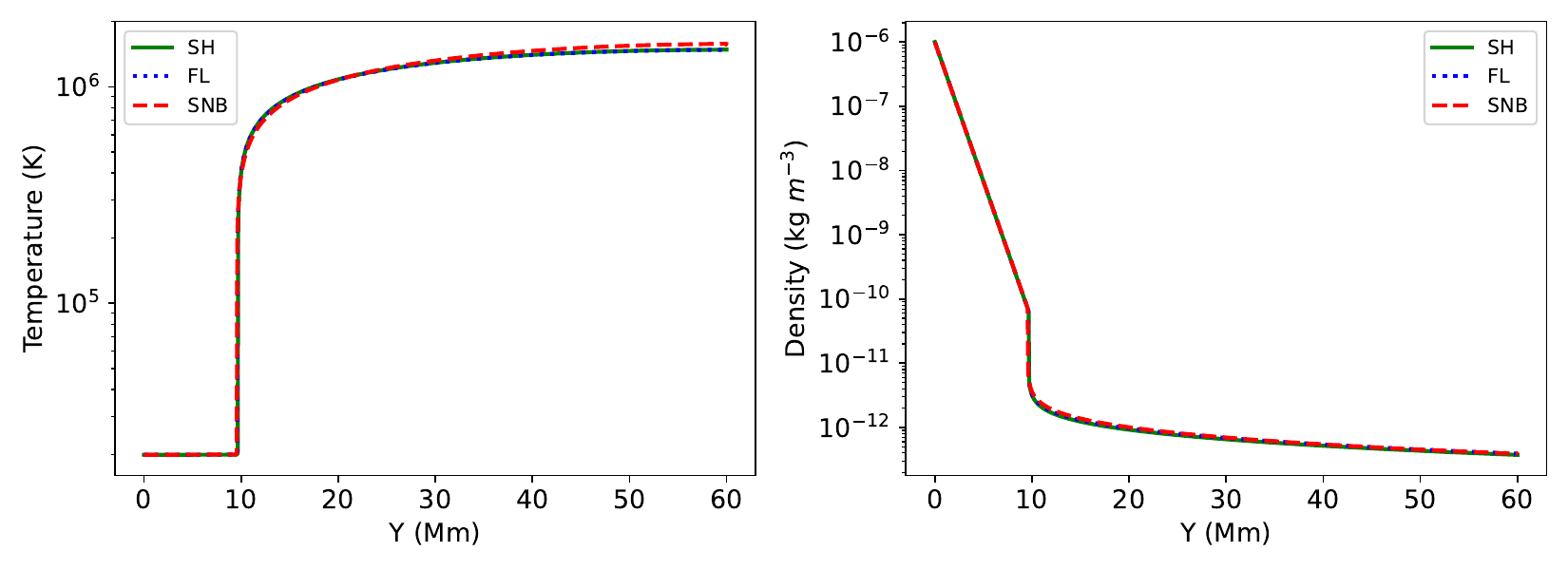}
\caption{The temperature (left) and density (right) of the 1d atmosphere used as the initial state of the simulation prior to magnetic reconnection and associated heating of the plasma. Note there are very slight differences between the equilibrium states between the different thermal conduction models due to weaker/stronger conduction across the transition region.}
\label{atmosphere}
\end{figure*}

Initially a 1-dimensional atmosphere is employed with a temperature profile given by 
\begin{equation*}
T(y) = T_{chr} + 0.5 \ (T_{cor} - T_{chr}) \left( \tanh \left(y-10\mathrm{Mm}\right) + 1\right),
\end{equation*}
where $T_{chr}$ is 20,000K and $T_{cor}$ is 1.2MK. Gravity is smoothly turned off near the top of the domain. $\rho$ is then calculated using hydrostatic equilibrium. Since this initial atmosphere is not in equilibrium once thermal conduction, radiation and heating are considered, it is evolved for approximately 25000s such that it reaches a stable state before any flare heating or magnetic field are introduced. The temperature and density of this atmosphere can be seen in Fig \ref{atmosphere}. The Transition Region Adaptive Conduction (TRAC) method outlined by \cite{Johnston2020} is used to increase thermal conduction and decrease radiative losses and background heating within the transition region to broaden unresolved length scales. TRAC has been shown to give accurate coronal temperature and density responses to heating for a grid resolution of 100km, while without it, a resolution of approximately 100m is needed \citep{Johnston2019}, which would not be computationally tractable for our simulations. Although the TRAC method was derived under the assumption of SH thermal conduction, simulations with FL and SNB show a similar temperature and density response. This can be attributed to a relatively low $K_n$ within the transition region, particularly after temperature gradients have been broadened by TRAC, and thus SH conduction remains valid in this region. As a result TRAC and the heat flux suppression of FL/SNB have limited interaction.

\subsection{Magnetic Field and Resistivity}

The initial state of the magnetic field is similar to numerous previous studies \citep{Yokoyama2001, Ruan2020} which is that of a vertical, force-free current sheet with components as follows:
\begin{equation}
B_x = 0, 
\end{equation}
\begin{equation}
B_y = B_0 \tanh(x / w),
\end{equation}
\begin{equation}
B_z = B_0 \ / \cosh(x/w),
\end{equation}
where $B_0$ ranges in value from 15G to 35G and $w$ is set to 5Mm.

To initiate magnetic reconnection and cause a rapid release of energy, a localised patch of high resistivity is placed near the centre of the domain:
\begin{equation}
\eta = 
\begin{cases}
\eta_0\left(1+\cos\left[\frac{\pi r}{w_{\eta}}\right]\right), & r \le w_{\eta} \\
0, & r > w_{\eta}
\end{cases}
,
\end{equation}
where $\eta_0$ is set to $1.12 \times 10^5 \ \mathrm{\Omega m} \ $(equivalent to a Lundquist number on the order of $10^7$), $r$ is $\sqrt{x^2 + (y-y_0)^2}$ with $y_0 = 30\mathrm{Mm}$, and $w_\eta$ = 2.4Mm. This resistivity is smoothly switched off at 45s to observe the dynamics post reconnection. Throughout the simulation, there is a uniform background resistivity of $1.12 \times 10^{-4} \  \mathrm{\Omega m}$ which smooths out grid-scale noise. 

Throughout this paper, we will focus on the transport and dynamics below this site of reconnection in order to study the chromospheric response. 

\section{Results}\label{sec:results}

\subsection{Temperature Evolution}

\begin{figure*}
\includegraphics[width = 180mm]{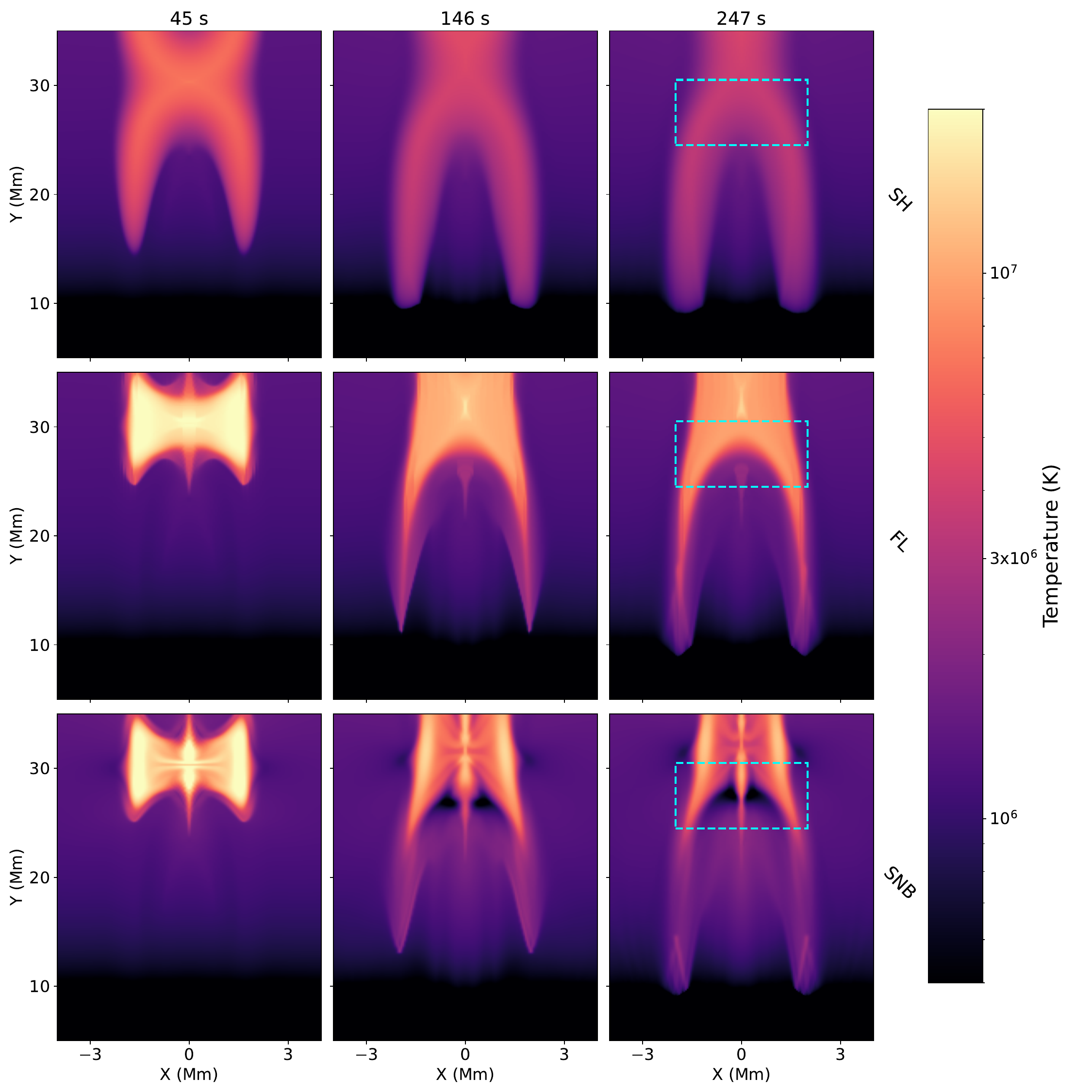}
\caption{Temperature distribution after magnetic reconnection is triggered with $B_0$ = 35G at three consequent times:  45, 146 and 247 seconds respectively. Each row corresponds to simulations carried out using a different thermal conduction model with SH (top), FL (middle) and SNB (bottom). Blue dashed boxes denote the regions used for averaging quantities over the loop apex. Note, the colour-scale is logarithmic.}
\label{2d-temp}
\end{figure*}

Figure \ref{2d-temp} displays the evolution of the plasma temperature over time following a resistive heating event using snapshots at 45s, 146s and 247s for each thermal conduction model. The 45s snapshots show the state at the point when anomalous resistivity is turned off and so at the termination of the heating. As can be seen in the top left panel of Figure \ref{2d-temp}, by this point a significant proportion of the energy released has been transported along the magnetic field and away from the reconnection site at $x=0, \ y=30 \ \mathrm{Mm}$ when using SH conduction, with conduction fronts close to approaching the transition region. However, for FL and SNB conduction, these fronts are still high up in the corona due to heat flux suppression slowing down thermal transport. This suppression keeps the released energy more localised, resulting in higher temperatures. It is noted that the total energy released due to magnetic reconnection is very weakly influenced by the thermal conduction model with a maximum difference of only 0.15\%. 

\begin{figure}
\includegraphics[width = 90mm]{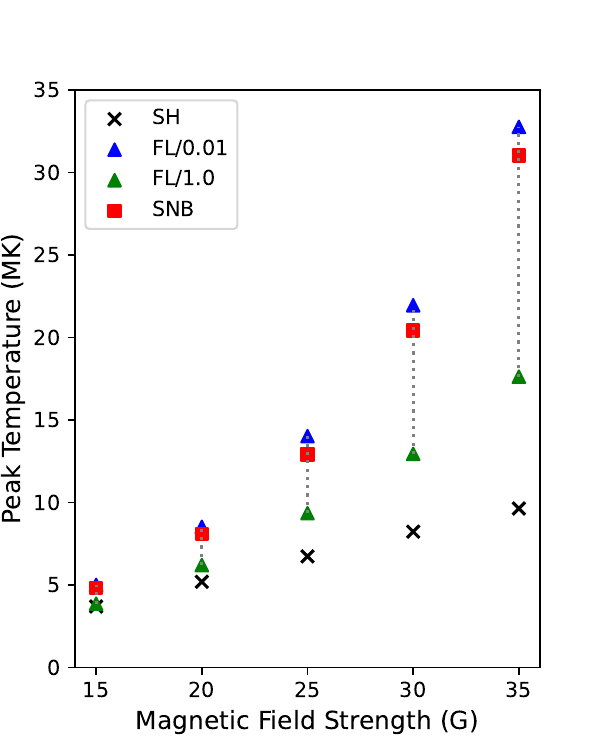}
\caption{The maximum temperature reached during the simulation for each conduction model and magnetic field strength used. Vertical dotted lines represent range of values obtained using different values for the flux-limiter.}
\label{Peak}
\end{figure}

Figure \ref{Peak} shows the difference in maximum temperatures reached between FL/SNB and SH transport. For the smallest flare with $B_0=15$G, which has a total energy release less than 20\% of the amount released during simulations with $B_0 = 35$G, peak temperatures are similar between all 3 thermal transport models. However, as the total energy released becomes larger, and so larger temperature gradients and reduced temperature length scales, non-local transport significantly increases the maximum temperature reached. Grey dotted lines between FL data points in Figure \ref{Peak} are used to show the range of values obtained with different flux-limiters $\alpha$. $\alpha = 1$ is the largest value used, and so has the weakest suppression of the heat flux. For $B_0=15$G, the Spitzer-H\"{a}rm heat flux only  slightly exceeds the free streaming limit and so $\alpha = 1$ gives a peak temperature within 0.1MK to that of SH. For $B_0 = 35$G however, even the large value of $\alpha=1$ gives strong suppression, resulting in higher temperatures. The simulations with the most strongly limited heat flux, using $\alpha=0.01$, show in all cases a noticeable increase in maximum temperature compared to $\alpha=1$ as reduced thermal conduction causes a higher energy density during the heating phase. For all magnetic field strengths, SNB lies within the range of temperatures from FL simulations, with $\alpha = 0.01$ giving the closest fit to the peak temperature out of the four $\alpha$ values used for all magnetic field strengths. However, as we will see later, $\alpha = 0.01$ does not provide the best fit to SNB for other observables such as downflow velocity or the timing of chromospheric evaporation.

In the 146s snapshots in Figure \ref{2d-temp}, a loop-like structure has been formed for all three conduction models. The temperature is more uniformly spread over this loop for SH, for which plasma has already reached the cool chromosphere, whereas the loop-top still contains higher temperatures for FL and SNB and thermal energy is still travelling downwards. 

Another feature of note is the spreading of high temperature plasma along the line $x = 0$, particularly visible in SNB panels of Figure \ref{2d-temp}, but also to a lesser extent in FL panels. When a stronger flux-limiter is used, such as $\alpha=0.01$, this effect becomes similar to that of SNB. This spreading can be attributed to hot plasma being advected along the reconnection jet. For SH conduction, the released energy is conducted away sufficiently quickly that the plasma in the reconnection jet is not sufficiently hot to be visible compared to the background. 

\begin{figure*}
\includegraphics[width = 180mm]{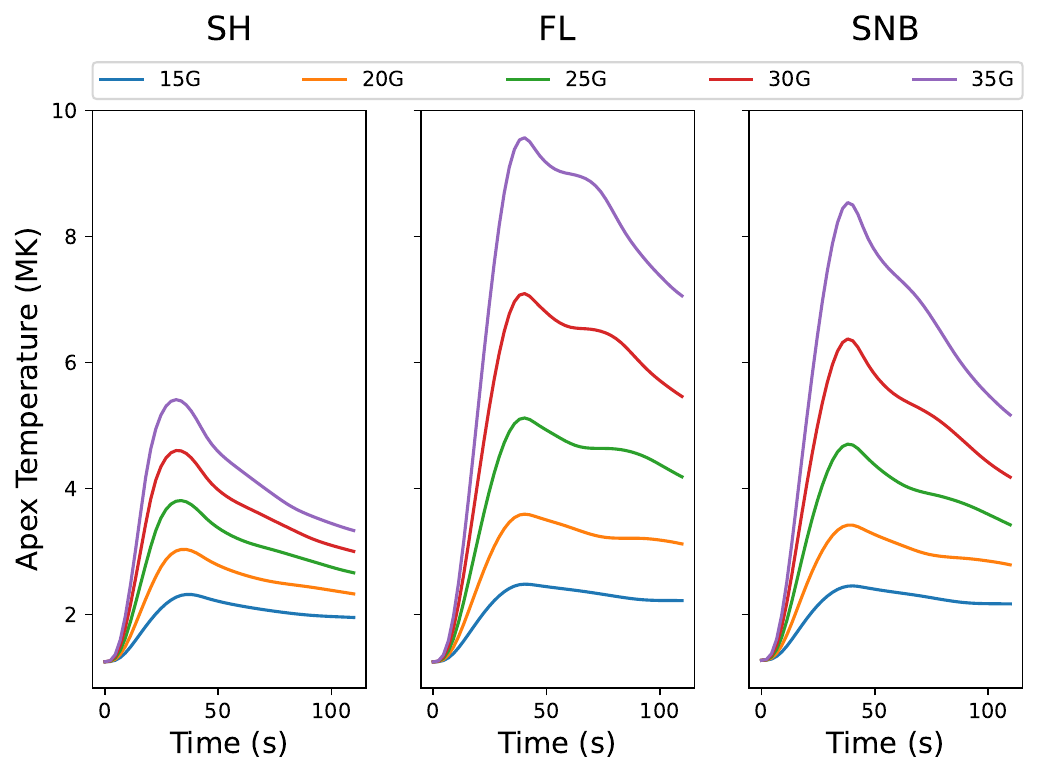}
\caption{The mean temperature over the apex region (shown in Figure \ref{2d-temp}) during the initial heating phase of the simulation for different values of $B_0$.}
\label{apexT}
\end{figure*}

At 247s, there is a noticeable difference in the temperature within the apex region, indicated with blue dashed boxes, between FL and SNB. Much of the extremely hot plasma has vacated this region for SNB where the hottest temperatures are found at the boundary of, and outside the apex region. For FL transport, the temperature is more uniform around the apex with the peak value found at the centre. Figure \ref{apexT} demonstrates this is also true in simulations with weaker magnetic field strengths. SNB consistently obtains a lower average temperature over the apex compared to FL, as well as a sharper decrease in temperature post the end of heating at 45s. It also reiterates that for smaller values of $B_0$, where less energy is released, the difference between SH and FL/SNB is greatly reduced.

\subsection{Density and Velocity Response}

\begin{figure*}
\includegraphics[width = 180mm]{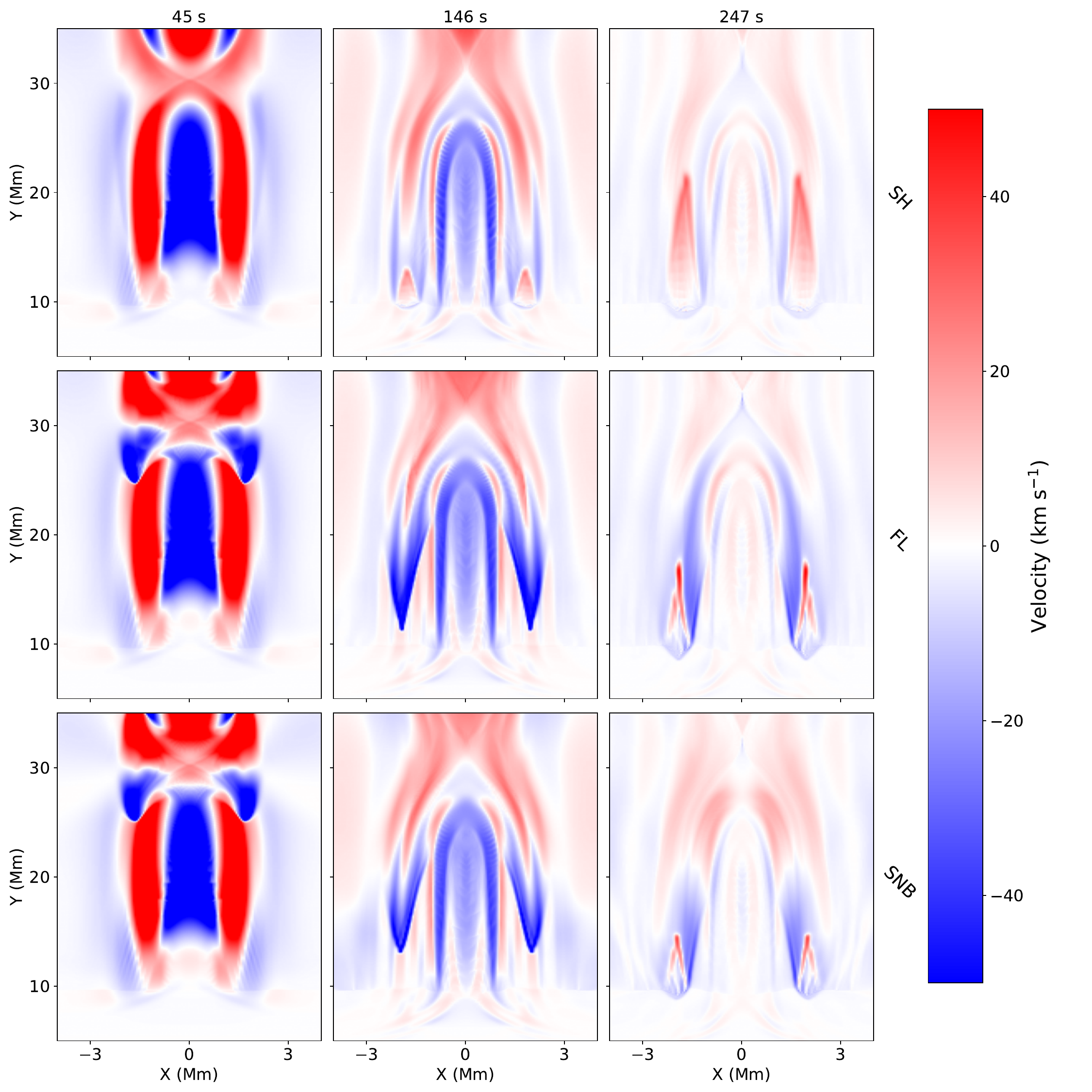}
\caption{Same as Figure \ref{2d-temp} but for the vertical velocity $v_y$. The colour scale here is linear with positive values (red) corresponding to upwards velocity.}
\label{2d-velocity}
\end{figure*}

\begin{figure*}
\includegraphics[width = 180mm]{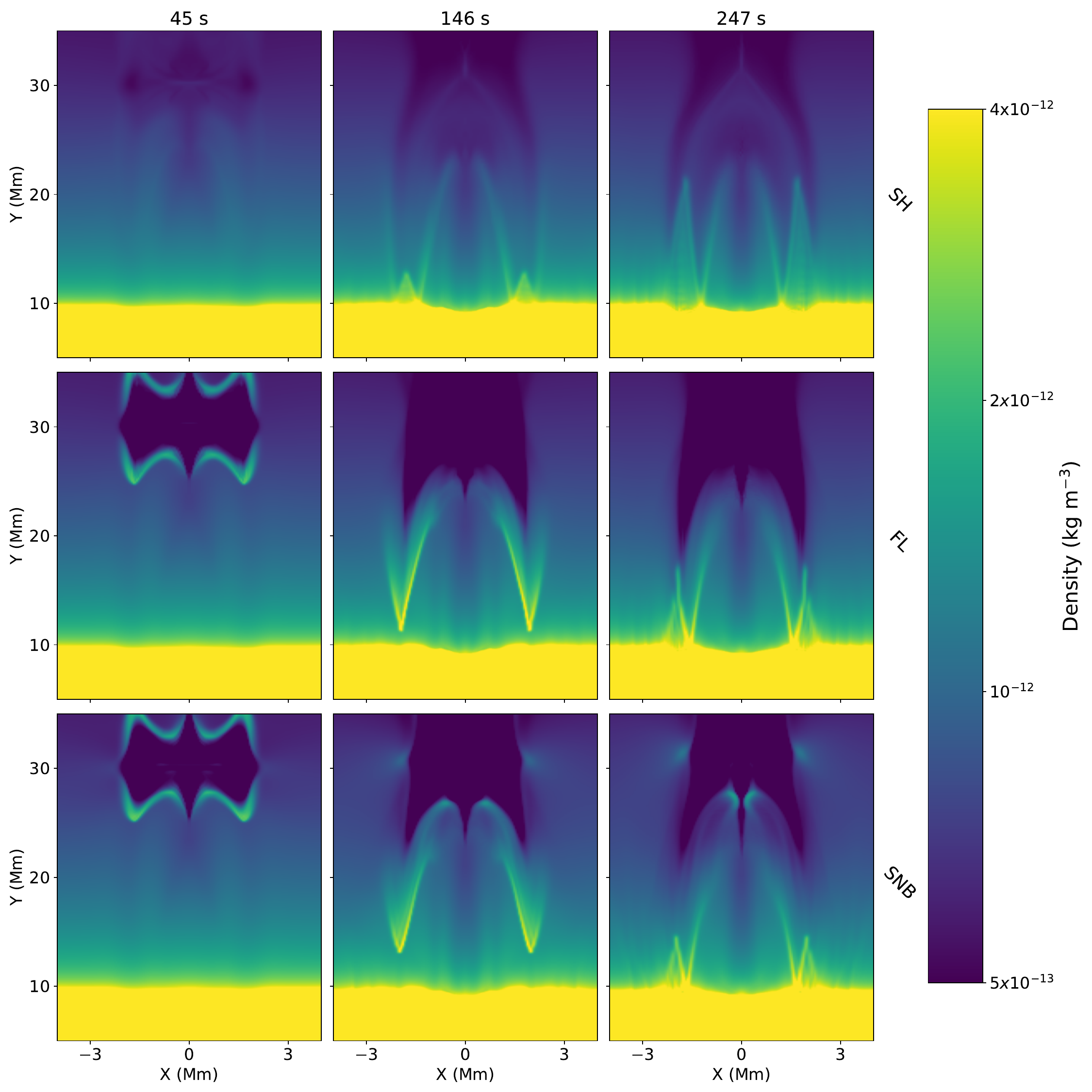}
\caption{Same as in Figure \ref{2d-temp} but for the mass density.}
\label{2d-density}
\end{figure*}

The rapid increase in temperature during heating provides steep pressure gradients which drive outflows from the reconnection site along the magnetic field which can be seen in the blue regions at approximately $x=\pm$ 2 Mm, $y=$ 25-30 Mm in 45s panels of Figure \ref{2d-velocity} for FL and SNB. The outflowing plasma then leads to a reduced density at and around the loop apex, as shown by dark region in Figure \ref{2d-density}. Due to the increased temperatures found in simulations using FL or SNB conduction, the apex sees a significant drop in density compared to SH, in addition to stronger velocity downflows. As well as these pressure driven flows, there are also vertical jets from the centre of magnetic reconnection moving along the line $x=0$. The velocities within these jets are similar in magnitude for all three conduction models, indicating that thermal conduction is not playing a significant role within reconnection. Due to the open boundary conditions at the top of the domain, there is significant outflow through this boundary from the reconnection jets which pulls some plasma from the below the reconnection site upwards as seen with in top of half of 146s panels.

\begin{figure}
\includegraphics[width=90mm]{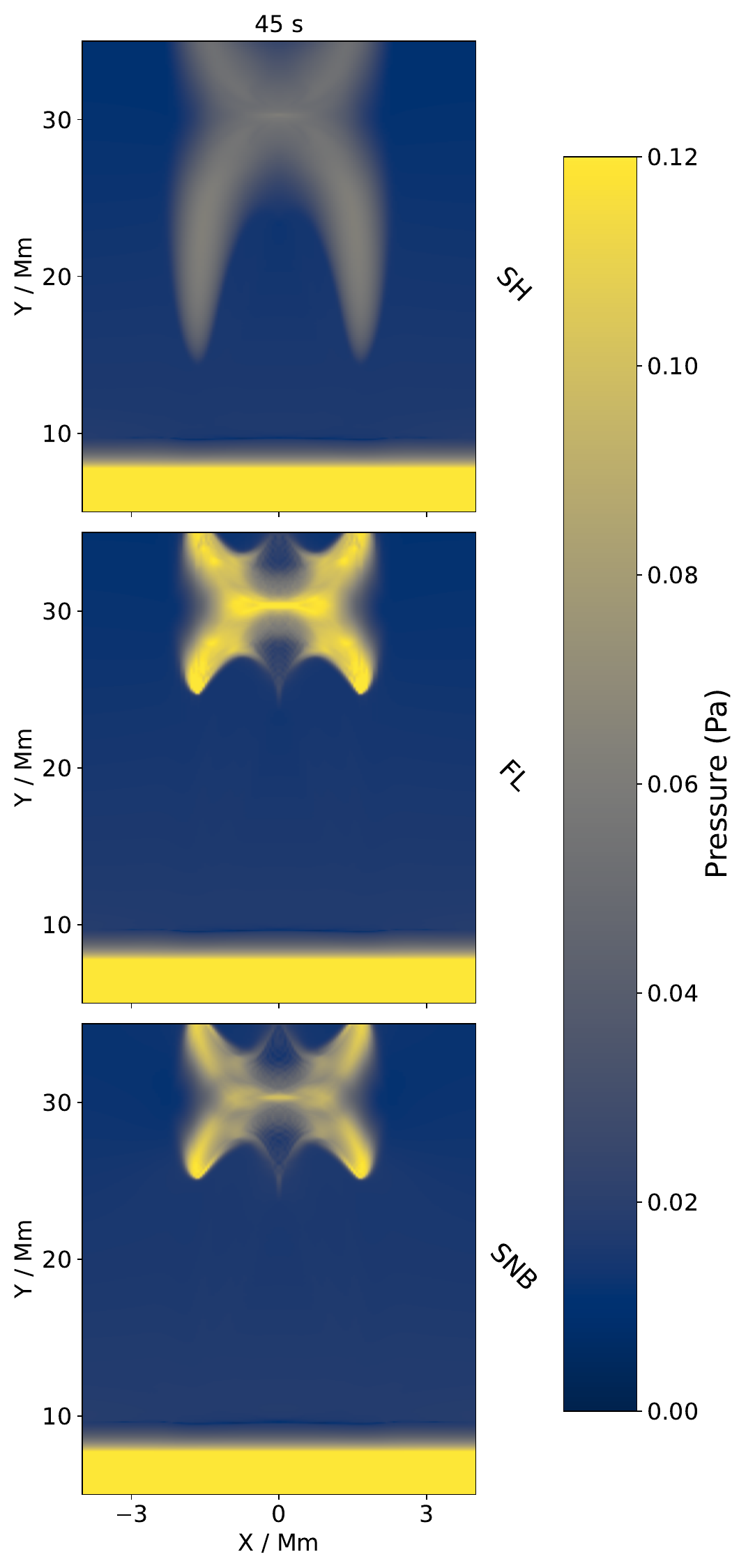}
\caption{The gas pressure at the moment when heating is switched off (45s after the start of heating) for SH (top), FL (middle) and SNB (bottom) conduction. Simulations use $B_0 = 35$G.}
\label{pressure}
\end{figure}

\begin{figure}
\includegraphics[width=90mm]{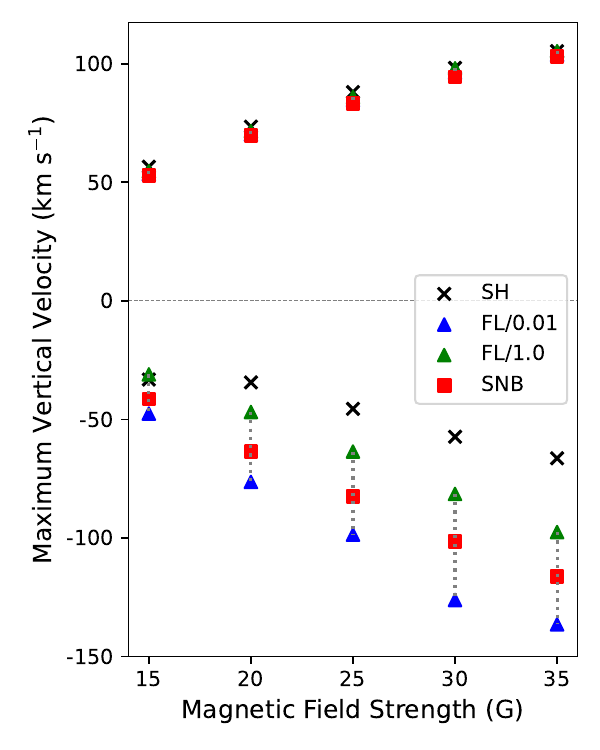}
\caption{The maximum vertical velocities $v_y$, positive denoting upwards and negative for downwards, reached in the footpoint region (as indicated by Figure \ref{2d-emission}) for each conduction model and magnetic strength used. As previously in Figure \ref{Peak}, vertical dotted lines between FL data point are used to display the spread of velocities found using different values of the flux-limiter.}
\label{velocity}
\end{figure}

The maximum downflow speeds from within the footpoint region are shown in the bottom half of Figure \ref{velocity}. The increased pressure and so pressure gradients found in the corona with non-local transport in the larger flares, as shown in Figure \mbox{\ref{pressure}}, lead to an increased downward velocity as thermal energy travels towards the chromosphere. As with the peak temperature, the range of velocities obtained using different flux-limiters is shown using grey dotted lines in Figure \ref{velocity}. Comparing across the magnetic field strengths $B_0 = 15, \, 25 \ \mathrm{and} \ 35$G, the best fits to SNB for the maximum downward velocity are $\alpha = 0.06, \, 0.2 \ \mathrm{and} \ 1.0$ respectively.

\subsection{Chromospheric Evaporation}

During solar flares and within the active solar corona, transient brightenings at extreme ultraviolet (EUV) wavelengths are observed corresponding to evaporation of heated chromospheric plasma \citep{Gupta2018}. As shown in Figures \ref{2d-velocity} and \ref{2d-density}, this process is seen throughout our simulations where as the hot downflows reach the chromosphere, dense plasma is ejected up into the corona. For SH, the more rapid thermal conduction means that these upflows can start to be seen at 146s originating from the top of the chromosphere at 10Mm while for FL and SNB, the plasma is still travelling down at this point. By 247s in the SH simulation the denser plasma has reached heights more than 10Mm above the transition region whereas FL and SNB are only just beginning to experience chromospheric evaporation with upward velocity at the loop footpoints. It is evident that the chromospheric evaporation here is generated as a result of thermal conduction with the initiation coinciding with hot plasma reaching the lower atmosphere. The preheating by fast electrons present in the SNB model appears not to give rise to any earlier chromospheric evaporation, indicating that energy transported this way does not provide rapid enough heating. The top half of Figure \ref{velocity} demonstrates that, while the downflow velocities are greater for the non-local models, the velocity at which chromospheric plasma is raised into the corona is not strongly influenced by the thermal conduction model used. However, there is an increase in upward velocity as the size of the flare is increased. 

\begin{figure*}
\includegraphics[width = 180mm]{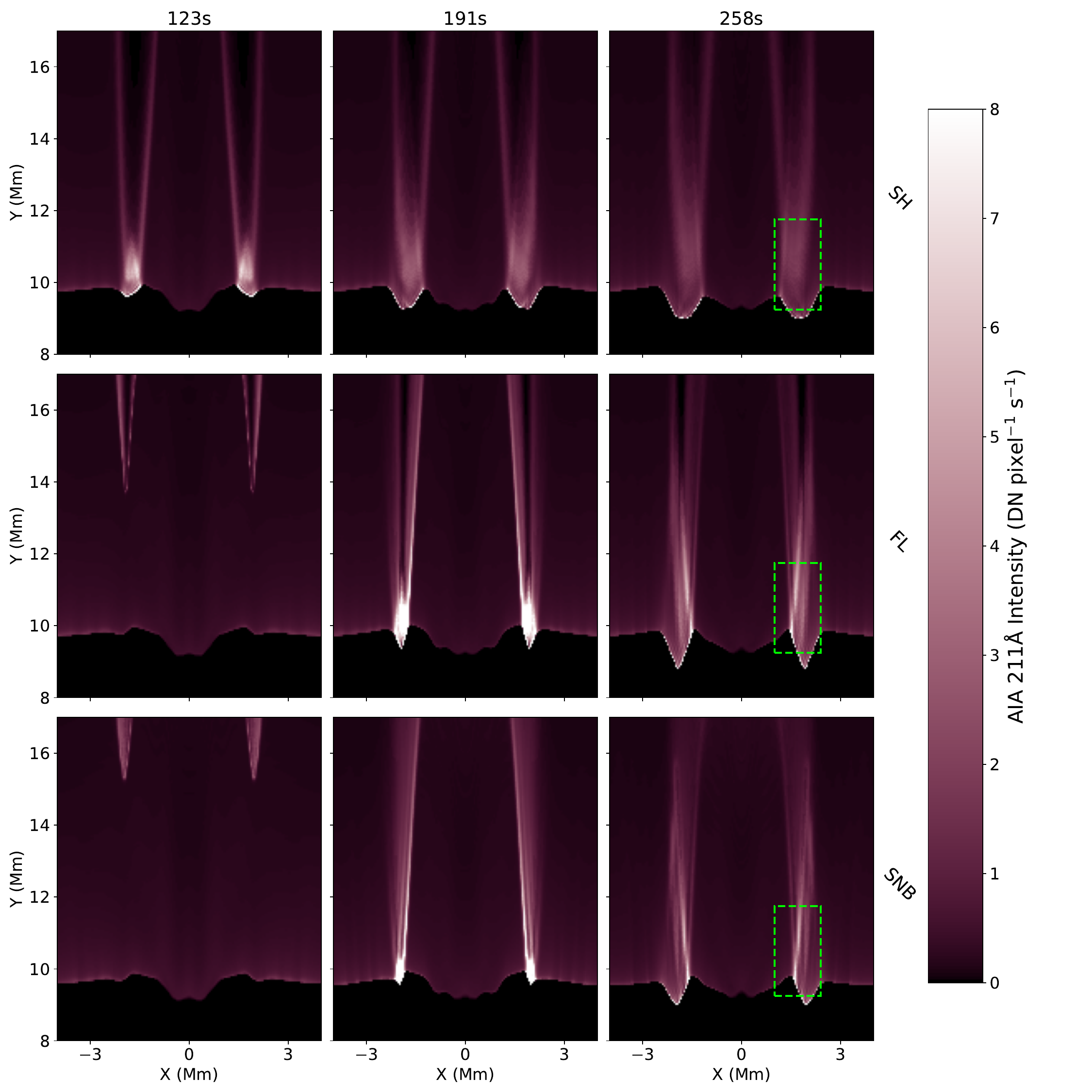}
\caption{Similar to Figures \ref{2d-temp}, \ref{2d-velocity} and \ref{2d-density} but for synthetic EUV emission corresponding to the 211\r{A} AIA band over a more zoomed in region lower in the atmosphere with snapshots taken at 123, 191 and 258s. Green dashed boxes represent the region used for averaging quantities over the loop footpoint.}
\label{2d-emission}
\end{figure*}

A series of synthetic images of the 211\r{A} AIA band is shown in Figure \ref{2d-emission} in the upper chromosphere and lower corona. These images were calculated using the SDO/AIA response functions from the FoMo database \citep{FoMo}, where a line of sight depth of 10Mm is assumed. Intensties are expressed in terms of Data Numbers (DN) the instrument would receive. The 211\r{A} band is chosen since the peak of its response function is at roughly 2MK, which coincides well with the evaporated plasma from our simulations. However, it is noted that our simulations also show bright footpoints at several other wavelengths, perhaps most significantly for the 94\r{A} and 171\r{A} passbands. At 123s, there is a significant brightening of the loop footpoint for the SH simulation. A similar brightening is then seen by 191s in FL and SNB images. By 258s, the emission has significantly reduced for all three conduction models, but still remains well above background levels.  

\begin{figure*}
\centering
\includegraphics[width=180mm]{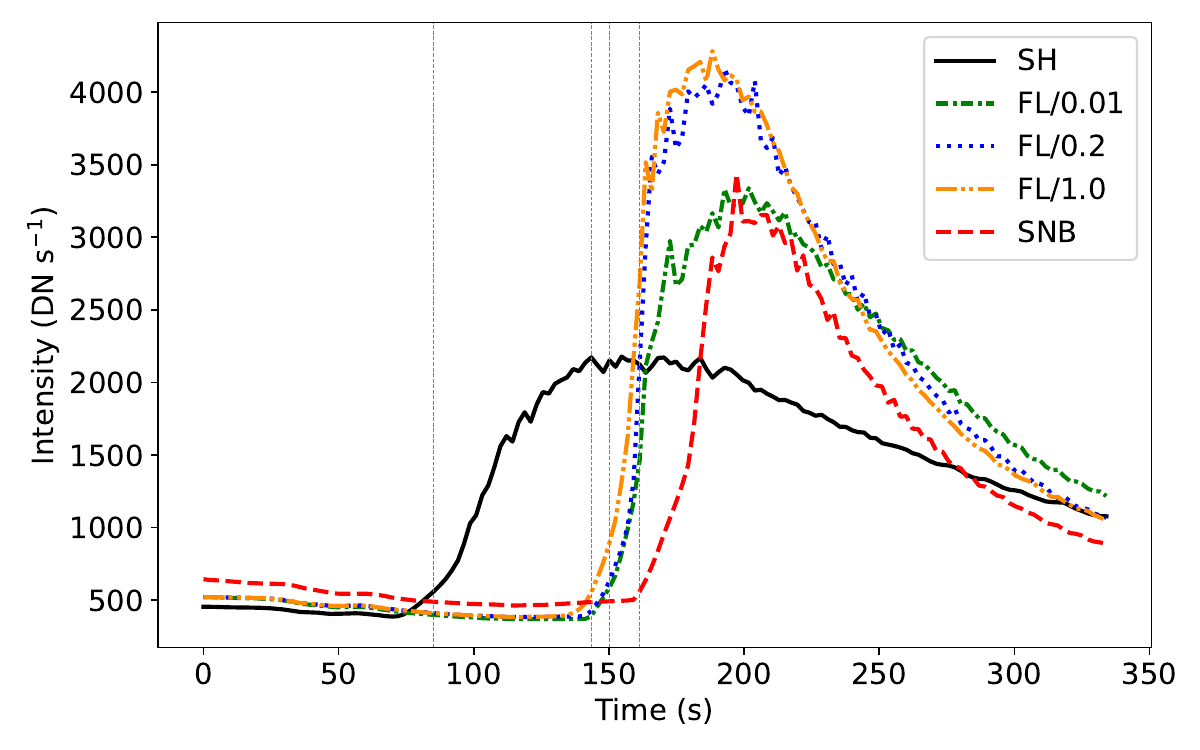}
\caption{Evolution of the EUV emission at 211\r{A} integrated over the footpoint of the resulting loop structure for simulations with $B_0$ = 35G (see Figure \ref{2d-emission} for exact region). To calculate an intensity a uniform depth of 10Mm is assumed in the z-direction. For the flux-limited (FL) simulations, the numbers 0.01, 0.2, and 1.0 correspond to the different values for the limiter $\alpha$. Vertical dotted lines are used to denote timings of when intensity begins to rise, indicating chromospheric evaporation has started.}
\label{footpoint}
\end{figure*}

The temporal evolution of the 211\r{A} emission is shown in more detail within Figure \ref{footpoint}. Vertical dotted lines indicate the timing of when the intensity first rises above a baseline of 550\,DN\,s$^{-1}$, and so give a delay time between the onset of magnetic reconnection and associated loop-top emission to the footpoint emission. These delay times are shown with $t_{rise}$ in Table \ref{table}. Both SNB and FL give significantly later emission than SH, with a higher maximum intensity. For $B_0=35$G, there is a 76s delay between SH and SNB EUV emission. This is balanced with a narrower peak for SNB with a FWHM of 63s compared to the 183s of SH. The FWHM of FL simulations lies in between these ranging from 83-114s. Table \ref{table} also gives the intensity ($I_{peak}$) of the maximum footpoint 211\r{A} emission for all parameters of magnetic field and flux-limiter. Across almost all magnetic field strengths, $t_{rise}$ appears consistently later with SNB than any of the FL values, which are within 7s for $B_0=35$G. The intensity of this peak also trends higher for FL than SNB, with both above that of SH conduction. From both Figure \ref{footpoint} and Table \ref{table}, it is interesting to note that the highest peak intensity for $B_0=35$G comes from the largest value of $\alpha$, which might be expected to be closest to SH, however this does not appear true for all magnetic field strengths. 

\begin{table*}
\caption{Timing and Intensity of Maximum AIA 211\,\AA\ Footpoint Emission}
\centering
\label{table:footpoint}
\begin{tabular}{l cc cc cc cc cc@{}}
\toprule
\toprule
& \multicolumn{10}{c}{Magnetic Field Strength $B_0$} \\
\cmidrule{2-11}
& \multicolumn{2}{c}{15 G} & \multicolumn{2}{c}{20 G} & \multicolumn{2}{c}{25 G} & \multicolumn{2}{c}{30 G} & \multicolumn{2}{c}{35 G} \\
\cmidrule{2-11}
& $t_{rise}$ & $I_{peak}$ & $t_{rise}$ & $I_{peak}$ & $t_{rise}$ & $I_{peak}$ & $t_{rise}$ & $I_{peak}$ & $t_{rise}$ & $I_{peak}$ \\
Model & (s) & (DN\,s$^{-1}$) & (s) & (DN\,s$^{-1}$) & (s) & (DN\,s$^{-1}$) & (s) & (DN\,s$^{-1}$) & (s) & (DN\,s$^{-1}$) \\
\midrule
SH                 & 309 & 583 & 197 & 766 & 152 & 1115 & 117 & 1633 & 85 & 2178 \\
FL ($\alpha=0.01$) & 327 & 640 & 226 & 957 & 188 & 1700 & 166 & 2665 & 150 & 3339 \\
FL ($\alpha=0.06$) & 314 & 675 & 226 & 1008 & 188 & 1716 & 164 & 2801 & 150 & 3565 \\
FL ($\alpha=0.2$)  & 325 & 698 & 226 & 1062 & 188 & 1775 & 164 & 3024 & 150 & 4163 \\
FL ($\alpha=1.0$)  & 321 & 675 & 220 & 990 & 184 & 1691 & 161 & 2826 & 143 & 4285 \\
SNB                & 316 & 703 & 229 & 937 & 199 & 1380 & 179 & 2113 & 161 & 3438 \\
\bottomrule
\end{tabular}
\par
\vspace{2ex}
{\raggedright \textbf{Notes.} $t_{rise}$ gives the timing of when the AIA 211\AA\ emission, integrated over the footpoint region displayed in Figure \ref{2d-emission} assuming a uniform line of sight depth of 10Mm, first rises above 550 DN\,s$^{-1}$. $I_{peak}$ gives the maximum intensity of this integrated emission. $t_{rise}$ and $I_{peak}$ are shown for each thermal conduction model and magnetic field strength used. Different values of $\alpha$ denote the different strengths of flux-limiter. Figure \ref{footpoint} gives the evolution of this emission over time for $B_0$ = 35G.\par}
\label{table}
\end{table*}

The top left panel of Figure \ref{energy} shows the upward mass flux from the regions of chromospheric evaporation found by integrating the positive component of $\rho v_y$ across the line $y=12$Mm, between $1.25 \, \mathrm{Mm}<x<2.5 \, \mathrm{Mm}$, assuming a uniform depth of 10Mm in the third dimension. There are some narrow peaks across all three conduction models during the first 100s or so which appear as a response to the reconnection jet. However, these are narrow and so while the flux is high at a specific point in time, they do not indicate large quantities of ejected material. At around 120s, the upward mass flux for SH begins to rise, followed by corresponding increases in the mass flux for FL and SNB. The timings of these increases line up well with the increases in EUV intensity shown in Figure \ref{footpoint}. Following the initial increase, the rate of chromospheric evaporation remains steady across all three models. The upward mass flux during chromospheric evaporation is similar across thermal conduction models which is in accordance with maximum upward velocity from Figure \ref{velocity}, with the data point for SH only slightly higher than SNB for $B_0 = 35$G.

\subsection{Energy Transport}

\begin{figure*}
\centering
\includegraphics[width = 180mm]{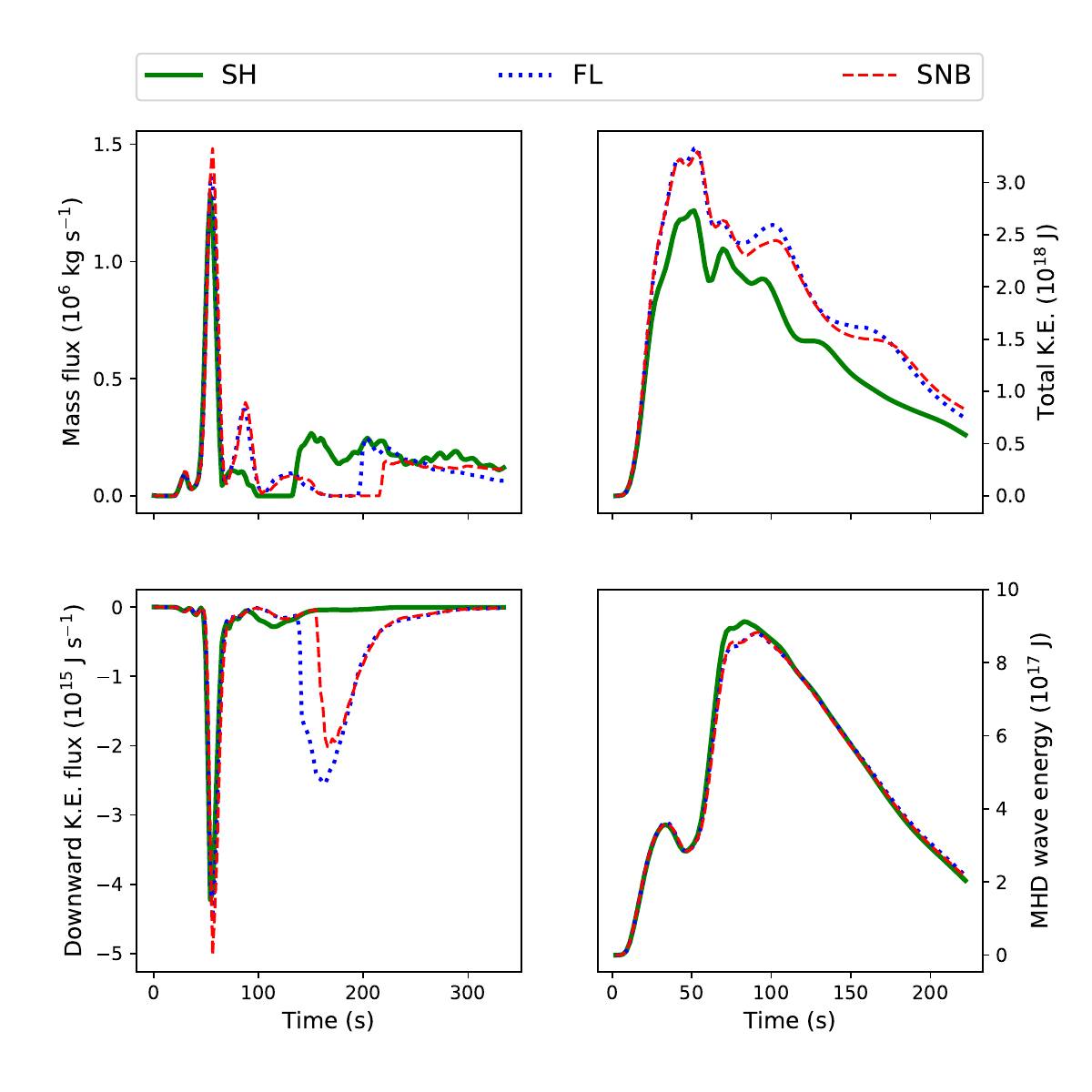}
\caption{Left column shows the upward mass flux (top) and downward kinetic energy flux (bottom) across the line y = 12Mm over time. The mass flux is only integrated for $x \in \left[1.25, 2.5\right]$ to better capture the evaporation region. Right column shows how the total kinetic energy (top) and MHD wave energy (bottom) integrated across the $x$ domain from $y=5\mathrm{Mm}$ to $y=30\mathrm{Mm}$ evolves over time. MHD wave energy is calculated by taking the kinetic energy perpendicular to the magnetic field. All four plots assume a uniform profile in the z-direction with a depth of 10Mm. Note the slightly shorter time period used for the right hand panels, in order to show the rise and fall phases more clearly.}
\label{energy}
\end{figure*}

The downward kinetic energy flux, as shown in bottom left panel of Figure \ref{energy}, is calculated by integrating $0.5 \rho v_y^3$ where it is negative over the line $y=12$Mm for each timestep. A depth of 10Mm is assumed in the third dimension. As seen earlier through the velocity snapshots in Figure \ref{2d-velocity}, there is an initial large downward kinetic energy flux from the reconnection jet which arrives in the lower atmosphere at $t\approx60$s. Then at a later time there are periods of downward kinetic energy flux corresponding to thermal conduction reaching the chromosphere. The magnitude of these secondary downward kinetic energy fluxes is in accordance with the maximum downward velocity seen in lower half of Figure \ref{velocity}, with largest flux for FL, then SNB and then a significant drop to SH which only shows a flux approximately 10 times smaller.

Quantifying the partition of released magnetic energy into different reservoirs is an important diagnostic of solar flares and provides a basis for comparison with observationally inferred flare energy budgets \citep{Aschwanden2014, Aschwanden2015}. The top right panel of Figure \ref{energy} shows how the total kinetic energy over the lower half of the domain ($5 <y< 30 \ \mathrm{Mm}$) varies over time. This is calculated by integrating the kinetic energy density $0.5 \rho v^2$ in space and assuming a depth in the third dimension of 10Mm. There is a sharp increase in kinetic energy around the time of the end of heating at 45s, where thermal energy is converted into kinetic energy through pressure gradients. FL and SNB transport give almost identical curves slightly above that of SH, which demonstrates a similar shape. Similarly, the bottom right panel of Figure \ref{energy} gives an estimate for the total fast magnetoacoustic wave energy calculated by integrating the component of the kinetic energy density perpendicular to the magnetic field over the same region. The rise in wave energy is delayed compared to that of kinetic energy, with the peak intensity 30s later. It is approximately 30\% of the magnitude of the kinetic energy. For reference, roughly 10\% of the total magnetic energy released is transferred into kinetic energy, meaning approximately 3\% of the total energy budget is carried as wave energy. The evolution of wave energy matches very closely across all three models, indicating it is driven directly from magnetic reconnection, and so there is minimal effect from thermal conduction.

\section{Conclusion and Discussion}\label{sec:conc}

We have carried out 2.5D MHD simulations of a localised resistive heating event within a region of the solar atmosphere in order to investigate the role of thermal conduction within solar flares and better understand the limitations of different thermal conduction models. Three different models were used: 1. The classical Spitzer-H\"{a}rm theory (SH) \citep{Spitzer1953}, 2. A local model where the heat flux is limited to a maximum value in terms of physical parameters (FL), and 3. The non-local model first presented by Schurtz, Nicola$\ddot{\i}$ and Busquet (SNB) \citep{Schurtz2000} which has been shown to have good agreement with the heat flux derived from VFP simulations in laser plasmas. Several different values for the flux-limiter are used to investigate whether there is a value that reproduces results from SNB in all cases, and how dependent the flare evolution is on the specific number chosen. Simulations are carried out under 5 different background magnetic field strengths $B_0$ ranging from 15G to 35G in order to give a range of flare sizes. 

In the quiet solar corona, purely local thermal conduction is demonstrated to be a good approximation to the physically better motivated SNB model in Figure \ref{atmosphere}, where there is only a minimal difference in temperature and density stratification between all three models.   

The evolution of various observables are different for flares when using either FL/SNB conduction compared to the SH simulations. The key differences are summarised as follows: 
\begin{enumerate}
\item{Non-local transport gives rise to greater peak and loop apex temperatures than SH as seen in Figures \ref{Peak} and \ref{apexT}. As the background magnetic field strength, and thus the total energy released, increases, this difference becomes more pronounced.}
\item{Transit time for thermal energy to move from the heating region down to the chromosphere is significantly slower with FL/SNB. This is most evident in Figure \ref{2d-temp}, with hot plasma reaching the lower atmosphere far earlier with SH. This leads to up to 76s later chromospheric evaporation under non-local transport and so a delayed increase in the corresponding EUV emission intensity.}
\item{Downflow velocities from the corona are larger when FL/SNB transport is used. This is due to the larger temperatures obtained during heating creating increased pressure gradients between the heated region and the lower corona. This is also seen when comparing the total kinetic energy across the domain with FL/SNB increasing beyond that of SH conduction after the initial heating phase. However, the subsequent upflow velocities are similar between all 3 models.}
\end{enumerate}
From these  discrepancies we can conclude that the use of classical Spitzer-H\"{a}rm conduction within MHD simulations of solar flares and active regions of the solar atmosphere is inaccurate and its use should be avoided if possible.

\par 

Simulations under FL and SNB transport are similar. For all magnetic field strengths, the maximum temperature reached with SNB conduction is within the range of those from different values of the flux-limiter. Figure \ref{velocity} shows that this is also true for the maximum downflow velocity. However, there are some differences between FL and SNB. While the peak temperature reached is similar between the two models, the hot plasma vacates the reconnection region faster with SNB, with temperature at the apex falling at a faster rate than even SH conduction. The apex is where the temperature gradients, and so $K_n$, are largest (reaching order 1), and so these conditions appear to be where the FL model breaks down and heating by suprathermal electrons only present in a truly non-local model becomes important. However for $K_n$ significantly above 1, the distribution function deviates sufficiently far from Maxwellian for the SNB model to also become invalid. These conditions could be reached if much more rapid / stronger heating were present, or if a large population of nonthermal electrons was accelerated during reconnection \mbox{\citep{Warmuth2020}}. Here nonthermal means departure of the electron
distribution function from Maxellian which is cannot be attributed by a temperature gradients. These nonthermal electrons can cause chromospheric evaporation and so would result in different timing and intensity to the conduction driven evaporation studied here. Studying such flares requires a kinetic approach, involving modelling across a large range of spatial scales, increasing the computational cost and so the feasibility for large scale multidimensional flare simulations.

The key drawback of the FL model is that it has a free parameter $\alpha$ which is usually set as a best fit to observations or other simulation results. We investigate this by using $\alpha \in \{0.01, 0.06, 0.2, 1\}$ and seeing if there is an optimal value. $\alpha=0.01$ gives the closest maximum temperature for all magnetic field strengths. A different outcome is seen for the maximum downflow velocity, where no single value gives the best fit and for the largest flare with $B_0=35$G, $\alpha=1$ is the closest to SNB. The intensity of EUV footpoint emission is lower in almost all cases for SNB than any value of the flux-limiter with no clear trend across all flare sizes when varying $\alpha$. The timing of footpoint EUV emission is similar between all $\alpha$ but consistently earlier than that obtained from SNB. However, more detailed spectroscopic diagnostics are required to constrain the plasma response to different thermal-transport models. In particular, the analysis of emission measure and Doppler shifts can directly probe flare-driven upflows and downflows, as demonstrated in \mbox{\citep{Battaglia2009,Imada2015}}, while density- and temperature-sensitive spectral lines could provide additional constraints on the thermodynamic evolution of the plasma. Future spectroscopic observations with the Multi-slit Solar Explorer (MUSE; \mbox{\citealt{DePontieu2019}}) and the Solar-C EUV High-Throughput Spectroscopic Telescope (EUVST; \mbox{\citealt{Shimizu2019}}) therefore have the potential to distinguish between the predicted plasma flows and temperature evolution produced by the different thermal-conduction models.

Therefore we conclude that while the flux-limiter is a significant improvement compared to Spitzer-H\"{a}rm conduction and the minimum requirement for solar flare simulations, there does not appear to be a flux-limiter value that works for all flare sizes or one that reproduces all elements of flare evolution best. Moreover, if, for example, a different background coronal atmosphere and density are used, this may again alter which value is best. Therefore, in order to achieve consistent and accurate multidimensional MHD simulations of the active solar corona, the SNB model currently provides the best practicable approach. If a full SNB treatment is too computationally expensive, a hybrid model where SNB is only switched on if the Knudsen number exceeds a certain threshold, and otherwise using a flux-limiter with $\alpha \approx 0.1$ could be a good compromise.

\begin{acknowledgments}
The work is funded by STFC Grant ST/X000915/1 and ST/Y509693/1. This work was performed using the DiRAC Data Intensive service at Leicester, operated by the University of Leicester IT Services, which forms part of the STFC DiRAC HPC Facility (www.dirac.ac.uk). The equipment was funded by BEIS capital funding via STFC capital grants ST/K000373/1 and ST/R002363/1 and STFC DiRAC Operations grant ST/R001014/1. DiRAC is part of the National e-Infrastructure.

\end{acknowledgments}

\section*{Data Availability} 
The simulation data and the code used to generate it are available from the corresponding author on reasonable request.

\software{Lare2d \citep{Arber2001} 
}

\bibliography{refs}{}
\bibliographystyle{aasjournalv7}

\end{document}